# On the formulas for effective stiffnesses in a paper by Zheng Ye, Victor L. Berdichevsky and Wenbin Yu

*Alexander A. Kolpakov, University of Neuchâtel, Switzerland*
*kolpakov.alexander@gmail.com*
**Alexander G. Kolpakov, SysAn, Russia*
*algk@ngs.ru*

A paper by Ye, Berdichevsky and Yu features a set of formulas (cf. formulas (11), (12), (19) in Ye et al. (2014)) for effective stiffnesses of corrugated plates. Without elaborating on said authors' actual approach leading to such formulas, we shall focus on the fact that they can be brought to a simpler form, especially if the corrugation wave is symmetric.

The most involved expression among (11), (12), (19) is the one for $A_{11}$, which is the effective stiffness of a given corrugated plate in the direction orthogonal to its corrugation profile (from here on we use the notation from Ye et al. (2014)). The most significant ingredient in computing $A_{11}$ is the following functional:

$$\langle \varphi A \rangle = \int_{-1/2}^{1/2} \varphi(Y) A(Y) dY, \qquad (1)$$

here $\varphi = \phi'$, while the function $Y = \phi(X)$ determines the corrugation profile in the dimensionless coordinates $(X,Y) = \varepsilon^{-1}(x,y)$, with $(x,y)$ being the standard Cartesian coordinates, and $\varepsilon$ being the length of the corrugation wave projection onto $Ox$ in $(x,y)$ coordinates (cf. Fig. 3 in Ye et al. (2014)). The above function $A(X)$ is defined by formula (13) in Ye et al. (2014):

$$A(X) = -\int_0^X \sqrt{a(Y)} \phi(Y) dY + B \int_0^X \sqrt{a(Y)} dY, \quad B = \frac{\langle \sqrt{a} \phi \rangle}{\langle \sqrt{a} \rangle}. \qquad (2)$$

(see also formula (5.68) in (Ye, 2013) since formula (12) for $B$ in Ye et al. (2014) has a typo).

We shall use integration by parts in order to transform (1) as shown below:

$$\int_{-1/2}^{1/2} \varphi(Y) A(Y) dY = \int_{-1/2}^{1/2} \phi'(Y) A(Y) dY = \phi(Y) A(Y) \Big|_{Y=-1/2}^{Y=1/2} - \int_{-1/2}^{1/2} \phi(Y) A'(Y) dY =$$
$$= -\int_{-1/2}^{1/2} \phi(Y) A'(Y) dY = \int_{-1/2}^{1/2} \sqrt{a(Y)} \phi^2(Y) dY - B \int_{-1/2}^{1/2} \sqrt{a(Y)} \phi(Y) dY \qquad (3)$$

In (3) we use the equality $A'(X) = -\sqrt{a(X)} \phi(X) + B\sqrt{a(X)}$, which is obtained by differentiating the first equality in (2), and the condition $\phi(\pm 1/2) = 0$, which follows from the definition of $\phi$, c.f. Fig.3 in Ye et al. (2014). Thus

$$\langle \varphi A \rangle = \langle \sqrt{a} \phi^2 \rangle - B \int_{-1/2}^{1/2} \sqrt{a(Y)} \phi(Y) dY. \qquad (4)$$

By plugging (4) into formulas (11), (12), (19) from Ye et al. (2014), the latter can be significantly simplified. These formulas reduce most when $B = 0$, which is the case e.g. if the corrugation wave is symmetric.

In $(X,Y)$ coordinates we use the dimensionless thickness of the corrugation profile $\tau = h/\varepsilon$ instead of its actual thickness $h$. The quantity $\tau$ determines if we can use the theory of thin plates (or beams) in our computations, cf. (Timoshenko and Woinowsky-Krieger, 2017).

We have that $\langle\sqrt{a}\phi^2\rangle = I_y / h\varepsilon^2$, where $I_y$ is the moment of inertia along the corrugation direction in $(x,y)$ coordinates (cf. formula (4) in Ye et al. (2014)), while the dimensionless functional $I_Y^0 = \langle\sqrt{a}\phi^2\rangle$, for which $(X,Y)$ coordinates are used, depends only on the corrugation profile. Thus $\langle\varphi A\rangle = I_Y^0$ in the case of $B = 0$.

As a result, in the case of a symmetric corrugation wave, when $B = 0$, by using the latter equality for $\langle\varphi A\rangle$ we can bring formula (19) for the effective stiffness $A_{11}$ from Ye et al. (2014) to the following compact form:

$$A_{11} = \frac{Eh}{1-\nu^2} \frac{1}{12\dfrac{\varepsilon^2}{h^2} I_Y^0 + \left\langle \dfrac{1}{\sqrt{a}} \right\rangle}$$

or

$$A_{11} = \frac{Eh}{1-\nu^2} \frac{1}{12 I_Y^0 \dfrac{1}{\tau^2} + \left\langle \dfrac{1}{\sqrt{a}} \right\rangle}, \quad \tau = \frac{h}{\varepsilon}.$$

The authors of the present note are grateful to the authors of Ye et al. (2014) for the discussion that helped improve the initial draft.